\documentclass[conference]{IEEEtran}
\IEEEoverridecommandlockouts
\usepackage{cite}
\usepackage{amsmath,amssymb,amsfonts}
\usepackage{graphicx}
\usepackage{textcomp}
\usepackage{xcolor}
\usepackage{booktabs}
\usepackage{multirow}
\usepackage{bm}       % For bold math symbols

\def\BibTeX{{\rm B\kern-.05em{\sc i\kern-.025em b}\kern-.08em
    T\kern-.1667em\lower.7ex\hbox{E}\kern-.125emX}}
\begin{document}

\title{Exploring Biomarker Relationships in Both Type 1 and Type 2 Diabetes Mellitus Through a Bayesian Network Analysis Approach
}

\author{\IEEEauthorblockN{Yuyang Sun}
\IEEEauthorblockA{\textit{Department of Engineering} \\
\textit{King's College London}\\
London, United Kingdom \\
yuyang.1.sun@kcl.ac.uk
}
\and
\IEEEauthorblockN{Jingyu Lei}
\IEEEauthorblockA{\textit{Division of Surgery and Interventional Science} \\
\textit{University College London}\\
London, United Kingdom \\
jingyu.lei.22@ucl.ac.uk}
\and
\IEEEauthorblockN{Panagiotis Kosmas}
\IEEEauthorblockA{\textit{Department of Engineering} \\
\textit{King's College London}\\
London, United Kingdom \\
panagiotis.kosmas@kcl.ac.uk}

}

\maketitle

\begin{abstract}
Understanding the complex relationships of biomarkers in diabetes is pivotal for advancing treatment strategies, a pressing need in diabetes research. This study applies Bayesian network structure learning to analyze the Shanghai Type 1 and Type 2 diabetes mellitus datasets, revealing complex relationships among key diabetes-related biomarkers. The constructed Bayesian network presented notable predictive accuracy, particularly for Type 2 diabetes mellitus, with root mean squared error (RMSE) of 18.23 mg/dL, as validated through leave-one-domain experiments and Clarke error grid analysis. This study not only elucidates the intricate dynamics of diabetes through a deeper understanding of biomarker interplay but also underscores the significant potential of integrating data-driven and knowledge-driven methodologies in the realm of personalized diabetes management. Such an approach paves the way for more custom and effective treatment strategies, marking a notable advancement in the field.
\end{abstract}
% In the field of diabetes research, understanding the complex relationships among various biomarkers is crucial for developing advanced treatment strategies. This study applies Bayesian network structure learning to analyze Type 1 and Type 2 diabetes mellitus datasets, revealing complex relationships among key diabetes-related biomarkers. Utilizing the Shanghai diabetes mellitus datasets, we constructed Bayesian networks that not only revealed significant relationships but also offered predictive insights, particularly for Type 2 diabetes mellitus. The constructed Bayesian network presented notable predictive accuracy, as validated through leave-one-domain experiments and Clarke error grid analysis. These findings provide a deeper understanding of diabetes dynamics and show the potential for combining data-driven and knowledge-driven methods in personalized diabetes management. 

\begin{IEEEkeywords}
Bayesian network, Structure learning, Glucose prediction, Diabetes management.
\end{IEEEkeywords}

\section{Introduction}
Diabetes mellitus (DM), a chronic metabolic disorder, has emerged as a global health crisis, affecting millions and escalating rapidly in prevalence and presenting significant challenges in diagnosis and management \cite{b6}. In addressing these challenges, continuous glucose monitoring (CGM) has been a pivotal development, offering real-time glucose data critical for effective DM management. While various methods exist for predicting glucose levels based on previous glucose trajectories \cite{b1,b2,b3,b4}, the analysis of diabetes-related biomarkers and their impact on glucose levels remains an area less explored, with many interrelationships yet to be fully understood \cite{b16,b17,b18}. Addressing this issue can therefore have a positive impact on the accuracy of prediction systems.

Recent advancements in machine learning, particularly in the realm of Bayesian networks (BNs) \cite{b7,b8}, present novel avenues to unravel the complex interplay between diabetes-related biomarkers and glucose measurements. BNs, known for their proficiency in handling uncertainties and probabilistic relationships, are suitable for modeling the complex interactions inherent in diabetes-related data. By building BNs on Type 1 and Type 2 DMs (T1DM and T2DM) datasets, we analyze the relationships between these biomarkers and glucose levels while considering the complexity and interdependencies of biomarkers. 

More specifically, our study leverages publicly available diabetes datasets \cite{b5}, applying Bayesian network structure learning \cite{b9,b10} to conduct a comprehensive analysis of key diabetes-related characteristics, including glycated hemoglobin (HbA1c), glycated albumin (GA), estimated glomerular filtration rate (eGFR), creatinine (CR), etc., and glucose measurements such as FPG and 2HPP. By categorizing the identified arcs as causal, correlated, or independent, the paper also uncovers relationships between these characteristics which can be both data and knowledge-driven, as discussed in our Results section.

% The implications of our findings enhance the field of diabetes research. Demonstrating the application of BNs in medical data analysis, this methodology also serves as a framework for researchers in biomedical engineering and related fields to explore complex relationships in their respective datasets. Consequently, this research not only aids in the selection of appropriate features for constructing prediction systems in diabetes research but also paves the way for broader applications, integrating data-driven and knowledge-driven decision-making for personalized strategy development across diverse scientific domains.

\section{Methodology}
This section outlines the methodology employed in our investigation of the interrelationships among diabetes-related biomarkers using BNs. Our approach encompasses the utilization of publicly available datasets \cite{b5} and sophisticated structure learning techniques to elucidate complex dependencies inherent in diabetes data. This methodology not only leverages advanced machine learning techniques but also tailors them specifically to address the unique challenges posed by the multifaceted nature of diabetes-related data.

\subsection{Shanghai Diabetes Mellitus Datasets}

Our study leverages the publicly available Shanghai DM datasets \cite{b5}, which include data on T1DM and T2DM. The ShanghaiT1DM dataset contains records from 12 T1DM patients, and the ShanghaiT2DM dataset includes data from 100 T2DM patients. These datasets record valuable anthropometric and biochemical characteristics alongside glucose measurements, forming the basis of our analysis.

The key features of these datasets include anthropometric characteristics such as age, weight, height, body mass index (BMI), and gender, and biochemical characteristics including HbA1c, GA, total cholesterol (TC), triglycerides (TG), high-density lipoprotein (HDL), low-density lipoprotein (LDL), CR, eGFR, uric acid (UA), and blood urea nitrogen (BUN). Additionally, glucose measurements, such as Fasting Plasma Glucose (FPG) and 2-Hour Postprandial Glucose (2HPP), are incorporated. These characteristics are categorized into five classes, as shown in Table~\ref{tab:my-table_1}.

\begin{table}[htb!]
\centering
\caption{Categorization of Diabetes-Related Characteristics in the Shanghai DM Datasets}
\label{tab:biomarkers-categorization}
\begin{tabular}{|l|l|}
\hline
\textbf{Categories} & \textbf{Characteristics}  \\
\hline
Glycemic biomarkers & HbA1c, GA \\
\hline
Anthropometric biomarkers & Age, Weight, Height, BMI, Gender \\
\hline
Lipid biomarkers & TC, TG, HDL, LDL \\
\hline
Kidney biomarkers & CR, eGFR, UA, BUN \\
\hline
Glucose measurements & FPG, 2HPP \\
\hline
\end{tabular}
\label{tab:my-table_1}
\end{table}

Detailed data specifications and additional features, including medical histories and complications, are available in the original dataset documentation \cite{b5}. In our analysis, we focus on characteristics that directly influence blood glucose fluctuations, such as factors related to glucose metabolism, lipid profiling, and kidney function. For data preprocessing, missing values were added using averages from other individual data when less than 20\% of data was missing for a given characteristic. Features with more than 20\% missing data, such as UA and BUN, were excluded from the Bayesian network structure learning process to maintain data integrity.

\subsection{Bayesian Network Structure Learning}

Our analysis utilizes BNs, probabilistic graphical models denoted as \( \mathcal{B} \) and defined by a tuple \( (G, \bm{\Theta}) \). Here, \( G \) represents a directed acyclic graph illustrating dependencies among random variables, while \( \bm{\Theta} \) encompasses parameters that define the strength of these dependencies. BNs are particularly valued in biomedical research for their interpretability, revealing complex dependencies that are often not straightforwardly causal, especially when latent variables are involved~\cite{b7,b8,b9,b11}.

In this study, we focused on structure learning of BNs to elucidate relationships among diabetes-related characteristics we mentioned before. Specifically, we employed the Tabu search algorithm \cite{b12}, complemented by Bootstrap re-sampling \cite{b13, b14} to generate two BNs respectively from the ShanghaiT1DM (12 samples) and ShanghaiT2DM (100 samples) datasets. The strength of arcs in generated BNs enables the identification of reliable dependencies by assessing the frequency of arc occurrence, with higher frequency indicating stronger dependencies. The robustness of these BNs was assessed by evaluating the frequency of arc occurrence in 100 Bootstrap iterations, with a strength threshold of $0.85$ for arc retention. The resulting 'Biomarkers\_Network' on the ShanghaiT2DM dataset combines arcs common to both DM models, arcs exceeding the strength threshold on T1DM, and some artificially added arcs for comprehensive analysis.

\begin{figure}[htb!]
  \centering
  \centerline{\includegraphics[scale = 0.12, trim = 0in 0in 0in 0in,clip]{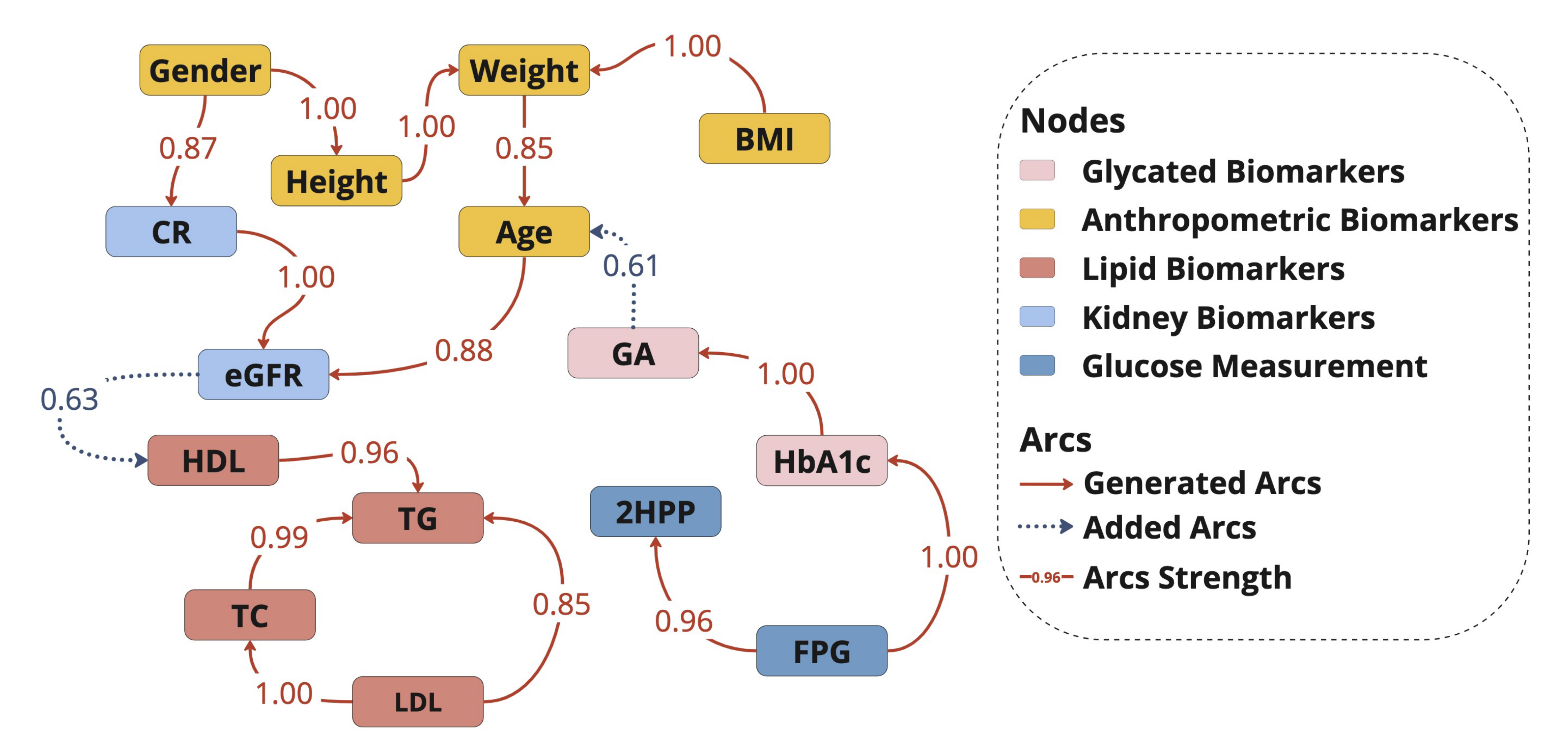}}
  \caption{Bayesian Network Structure for Diabetes Biomarker Analysis on both ShanghaiT1DM and ShanghaiT2DM Datasets.}
\label{fig:fig_label_1}
\end{figure}

'Biomarkers\_Network', as depicted in Figure~\ref{fig:fig_label_1}, visualizes these dependencies through its structure of arcs and nodes. The nodes are categorized into four biomarker types and one measurement category, as tabulated in Table \ref{tab:my-table_1}). The arcs indicate statistical correlations between connected variables. The 'arc strength', depicted at the arcs, quantifies the probability and directionality of these dependencies.

Notably, the main differences in the generated BNs between T1DM and T2DM are mainly due to the sample size., with T2DM's larger dataset revealing more potential arcs. Some arcs, while not meeting the high strength threshold, were artificially introduced in 'Biomarkers\_Network' to ensure a full connection of all available nodes to maximize the consideration of all variables for our dependencies analysis and prediction experiments. Further discussion on correlation and causality analysis is presented in the subsequent discussion section.

Performance evaluation of 'Biomarkers\_Network' was conducted through leave-one-domain experiments, setting aside one individual's data for testing while using the remaining individuals' data for network training. The prediction accuracy for FPG and 2HPP values was evaluated using mean absolute error (MAE) and root mean squared error (RMSE), evaluated at each individual, and then calculated as the averages. Additionally, a Clarke error grid \cite{b15} is introduced to visualize the predicted results of T1DM and T2DM.

\section{Results and Discussion}
\subsection{Prediction Results of Leave-One-Domain Experiment
}
The performance of 'Biomarkers\_Network' in predicting glucose levels was evaluated using leave-one-domain experiments. These experiments were conducted separately for T1DM and T2DM datasets to predict FPG and 2HPP values, employing MAE and RMSE as evaluation metrics. The raw FPG levels for T1DM range from 117.00 to 262.35 mg/dL, whereas for T2DM they span from 126.00 to 194.40 mg/dL. Regarding the raw 2HPP levels, T1DM exhibits a range of 248.76 to 348.84 mg/dL, compared to 196.16 to 317.88 mg/dL for T2DM. The predicted results, as detailed in Table~\ref{tab:my_label_2}, demonstrate the network's capability in predicting glucose levels, with a notably better performance observed in the T2DM dataset compared to T1DM. Notably, the superior performance in the T2DM dataset is attributed to its larger size and the complexity of interactions it captures. The results highlight the potential of Bayesian networks in analyzing and predicting key diabetes metrics, laying the groundwork for further research and application in diabetes management.

\begin{table}[htb]
    \caption{Leave-one-domain experiment results (Mean) of 'Biomarkers\_Network'.}
    \centering
    \begin{tabular}{c|c|c|c}
        \hline
        \textbf{Datasets} & \textbf{Label} & \textbf{MAE (mg/dL)} & \textbf{RMSE (mg/dL)}  \\
        % \multirow{2}{*}{\textbf{Datasets}} & \multirow{2}{*}{\textbf{Label}} & \textbf{MAE} & \textbf{RMSE}  \\
        % & & \textbf{(mg/dL)} & \textbf{(mg/dL)}  \\
        
        \hline
        T1DM & FPG  & 30.29 & 36.16  \\
             & 2HPP & 31.94 & 41.59  \\
             \hline
        T2DM & FPG  & 19.22 & 28.23  \\
             & 2HPP & 28.99 & 40.12 \\
        \hline
    \end{tabular}
    \label{tab:my_label_2}
\end{table}

\subsection{Clarke Error Grid visualization}

\begin{figure}[htb!]
  \centering
  \centerline{\includegraphics[scale = 0.55, trim = 0in 0in 0in 0in,clip]{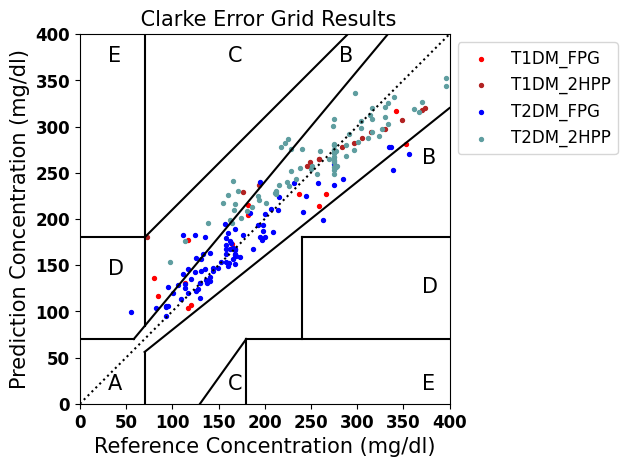}}
  \caption{Clarke Error Grid Analysis for T1DM and T2DM Predictions. FPG and 2HPP predictions for T1DM (red and brown) and T2DM (blue and cyan) are displayed across the grid.}
\label{fig:fig_label_2}
\end{figure}

For visualization, Clarke error grid analysis \cite{b15}, illustrated in Figure \ref{fig:fig_label_2}, is introduced to provide insights into the model's predictive accuracy. The grid categorizes results into zones reflecting clinical impact: Zone A (clinically acceptable), Zone B (benign errors), and Zones C to E (errors with potential clinical significance). The T1DM predictions are predominantly within Zones A and B, denoting acceptable accuracy. 2HPP predictions show a slightly higher dispersion, suggesting increased variability in postprandial glucose responses. Conversely, T2DM results are more concentrated in Zone A for both FPG and 2HPP, denoting enhanced reliability and clinical applicability. This visual analysis corroborates the numerical findings, where T2DM showed lower MAE and RMSE values compared to T1DM, and underscores the model's clinical viability for diabetes management.

\subsection{Analysis of Relationships in the Bayesian Network}
The structure of the Bayesian network, derived through a data-driven approach, reveals intricate relationships among diabetes-related biomarkers. This subsection offers a knowledge-driven approach, utilizing expert knowledge to interpret and validate these relationships.

\subsubsection{Glucose Measurement and Glycated Metrics Relationships}
\begin{itemize}
    \item \textbf{'FPG' to 'HbA1c' Arc:} 
    The strong arc strength (1.0) observed between FPG and HbA1c in our network aligns with established medical knowledge. FPG is a direct measure of blood glucose after fasting, while HbA1c offers a long-term glycemic index. While high FPG levels over time contribute to increased HbA1c, the relationship isn't strictly causal, considering HbA1c's sensitivity to various other factors beyond fasting glucose levels. This \textbf{correlation} is well-supported by several studies \cite{b19,b20}, highlighting the interdependence of these metrics in diabetes management. This arc in our network reflects this well-established interdependence, crucial for diabetes diagnosis and management.
    % This arc, with a strength of 1.0 for both BNs from T1DM and T2DM, illustrates a significant correlation between FPG levels and HbA1c values. FPG, a critical biomarker for diabetes diagnosis, reflects the concentration of glucose in the blood after fasting. HbA1c, on the other hand, provides an overview of long-term blood glucose control by measuring the percentage of glucose-attached hemoglobin. While high FPG levels over time contribute to increased HbA1c, the relationship isn't strictly causal, considering HbA1c's sensitivity to various other factors beyond fasting glucose levels. This correlation is well-supported by several studies \cite{b19,b20}, highlighting the interdependence of these metrics in diabetes management.
    
    \item \textbf{'HbA1c' to 'GA' Arc:} Another notable arc is between HbA1c and GA, with strength values of 0.96 and 1.0 across our models. The positive \textbf{correlation} between these two metrics is reinforced by existing literature \cite{b23,b24}. GA, unlike HbA1c, offers a shorter-term view of glucose control, making it especially useful in cases where HbA1c results might be unreliable. However, establishing a causal link is challenging due to the distinct periods these metrics reflect.
    
    \item \textbf{'FPG' to '2HPP' Arc:} 
    % The difference in arc strength in BN of T2DM (0.96) compared to BN of T1DM (0.35) is clear. It highlights the pathophysiological differences between T1DM and T2DM \cite{b25}, as well as the impact of varied treatment regimens on glucose variability. This finding emphasizes the need for a nuanced approach in managing these two types of diabetes, considering the inherent differences in their glucose control mechanisms.
    The difference in the arc strength between T1DM (0.35) and T2DM (0.96) networks highlights the distinct pathophysiological profiles of these conditions. T2DM's higher \textbf{correlation} suggests a stronger link between fasting and postprandial glucose levels, a phenomenon well-documented in diabetes research \cite{b25}. This differential relationship underscores the need for distinct management strategies for T1DM and T2DM.
\end{itemize}

\subsubsection{Anthropometric Metrics Relationships}
\begin{itemize}
    \item \textbf{'Gender' to 'Height' Arc:} The relationship between gender and height in our network is a clear \textbf{causal} example of genetic and hormonal influences on physical characteristics, extensively corroborated by global health data \cite{b26}. The presence of this arc is a validation of the network's ability to capture existing fundamental biological relationships by a data-driven approach.
    
    \item \textbf{Arcs among 'Height', 'Weight', and 'BMI':} The \textbf{correlations} among these anthropometric measures are well-established in physiological research \cite{b27}. BMI, calculated using height and weight, serves as a key health indicator. The strength of these arcs, while significant, varies due to the dataset sizes, underscoring the importance of considering data variability in such analyses.
\end{itemize}

\subsubsection{Lipid Metrics Relationships}
These lipid metrics are all critical in assessing cardiovascular risk in diabetes patients. The relationships are categorized as \textbf{correlations} due to the complex interplay of metabolic and biochemical factors influencing these metrics\cite{b28,b29,b30}. Notably, the Friedewald Equation \cite{b28} demonstrates these dependencies, though pinpointing direct causal links remains a challenge.

\subsubsection{Kidney Metrics Relationships}
The network underscores the relationship between serum CR levels and eGFR, as well as the influence of age on eGFR. These \textbf{correlations} align with nephrology research \cite{b31,b32,b33,b34,b35}, which emphasizes the importance of these biomarkers in assessing kidney health in diabetes patients. Besides, the MDRD equation \cite{b31,b32} provides a quantitative framework, illustrating how CR levels and age are crucial in estimating eGFR, with gender and ethnicity as additional factors.

\subsection{Limitations and Future Directions}

Our Bayesian network has provided valuable insights into the relationships among diabetes-related biomarkers. However, it's important to note that deducing causality from such models is inherently challenging. It often necessitates access to more extensive datasets and more refined modeling techniques capable of handling complex biological interactions. Future research directions include expanding the datasets in both size and diversity to enhance model robustness and incorporating patient data from real-world settings. This expansion would allow for the exploration of dynamic models that better capture the temporal fluctuations of diabetes biomarkers. Additionally, the application of these analytical insights to develop decision-support tools or to tailor personalized diabetes management strategies holds great promise. Such advancements have the potential to transform patient care and improve clinical outcomes in diabetes management.

\section{Conclusion}
In this study, we applied Bayesian network structure learning to T1DM and T2DM datasets, uncovering the intricate relationships among various diabetes-related biomarkers. This analysis has deepened our understanding of the complex interplay of factors involved in diabetes, thereby enriching the broader knowledge base in this area. The notable predictive accuracy observed, especially within the T2DM dataset, underscores the practical utility of our methodology in diabetes management. Our future work will explore further this methodology of combining knowledge and data-driven approaches in biomedical research.

\end{document}